\pdfoutput=1
\documentclass[a4paper,11pt]{article}

\usepackage{fullpage}
\usepackage{amsmath,amsfonts}
\usepackage{hyperref}
\usepackage{graphicx,subfigure}
\newcommand{\MMA}{\emph{Mathematica}}
\newcommand{\RandFile}{\texttt{RandFile}}

\newcommand{\SubList}[2]{\ensuremath{\{#1_1,#1_2,\dots,#1_#2\}}}

\usepackage{listings} % code formatter
\usepackage{courier}
\usepackage{caption} % ttfamily font for code formatter

\newcommand{\proglang}[1]{\textbf{#1}}
\newcommand{\pkg}[1]{\textbf{#1}}
\newcommand{\ie}{\emph{ie.}}

\title{\RandFile\ package for \MMA\ for accessing file-based sources of randomness}
\author{J.A Miszczak${}^{1,}\footnote{E-mail: miszczak@iitis.pl}$ \quad M. Wahl${}^{2}$\\
${}^{1}$Institute of Theoretical and Applied Informatics,
Polish Academy of Sciences,\\
Ba{\l}tycka 5, 44-100 Gliwice, Poland\\
${}^{2}$PicoQuant GmbH,\\ Rudower Chaussee 29 (IGZ),
12489 Berlin, Germany
}

\date{15/03/2015 (v. 0.30)}

\begin{document}
\maketitle

\begin{flushright}
\begin{minipage}{0.42\textwidth}
\emph{\small After 40 years of development, one might think that the making of random
numbers would be a mature and trouble-free technology, but it seems the creation
of unpredictability is ever unpredictable.}
\end{minipage}\\[10pt]
\small B. Hayes~\cite{hayes93wheel}
\end{flushright}

\begin{abstract}
We present a package for \MMA\ computer algebra system which allows the
exploitation of local files as sources of random data. We provide the
description of the package and illustrate its usage by showing some examples. We
also compare the provided functionality with alternative sources of randomness,
namely a built-in pseudo-random generator and the package for accessing hardware
true random number generators.
\end{abstract}

%\tableofcontents

%%%%%%%%%%%%%%%%%%%%%%%%%%%%%%%%%%%%%%%%%%%%%%%%%%%%%%%%%%%%%%%%%%%%%%%%%%%%%%%%
\section{Introduction}\label{sec:intro}
%%%%%%%%%%%%%%%%%%%%%%%%%%%%%%%%%%%%%%%%%%%%%%%%%%%%%%%%%%%%%%%%%%%%%%%%%%%%%%%%
Random numbers are essential for solving a variety of physical and mathematical
problems \cite{fishman,metropolis87beginning} and the quality of random numbers
used in simulations is critical for the
outcomes~\cite{ferrenberg92montecarlo,bauke04pseudo}. It is of no surprise that
the considerable research effort has been devoted to the development of methods
for generating pseudo-random \cite{park88random,knuth-vol2} and truly random
numbers \cite{hotbits,random.org,kanter10optical}.

One of the main advantages of pseudo-random number generators is their ability
to reproduce the sequence of random numbers used in a given simulation. However,
one can argue wether the amount of randomness provided by such generators is
sufficient~\cite{ferrenberg92montecarlo,bauke04pseudo}. On the other hand,
random numbers generated using hardware generators (e.g. quantum random number
generators \cite{wahl11ultrafast,quantis-support}) provide fast sources of
truly random numbers. In this case, however, the user has to have a specific
hardware at their disposal and loses the ability to reproduce the sequence of
random numbers used during the simulation. The ideal solution would allow the
user to employ true random data sources without the need for a specific hardware
and with the ability to repeat the experiments.

The presented work aims to fill this gap. We provide
\RandFile~\cite{randfile-www} package for \MMA\ computer
algebra system, which allows using files with random data for generating
random numbers used in simulations~\cite{comqm-notes} easily. There has been
twofold motivation for the work on the provided package. Firstly, the presented
package allows using fast, high-quality sources of randomness in simulations.
Secondly, as with the help of the presented package one can use any file as a
source of data for producing random numbers, this enables users to observe the
influence of the quality of random data on the outcomes of conducted
simulations.

This paper is organized as follows.
In Section~\ref{sec:rand-file} we provide an overview of the presented package
and describe its functionality.
In Section~\ref{sec:comparison} we compare the presented package with the
alternative methods providing sources of randomness available for \MMA\ users.
In Section~\ref{sec:examples} we provide some usage examples.
Finally, in Section~\ref{sec:final} we provide the summary of the presented work
and some concluding remarks.

%%%%%%%%%%%%%%%%%%%%%%%%%%%%%%%%%%%%%%%%%%%%%%%%%%%%%%%%%%%%%%%%%%%%%%%%%%%%%%%%
\section{Overview of the package}\label{sec:rand-file}
%%%%%%%%%%%%%%%%%%%%%%%%%%%%%%%%%%%%%%%%%%%%%%%%%%%%%%%%%%%%%%%%%%%%%%%%%%%%%%%%
Package \RandFile\ is designed to provide a user interface for accessing
random numbers generated using random binary data stored in files on a hard
drive. In the following we assume that the user has obtained a sample of random data
which are stored in \lstinline{sample_data.bin} file and accessible for the
\MMA\ system. Such file, containing 1MB of random data is
distributed with the package. 

One should note that the \lstinline{.bin} file extension is used here to
indicate that the file contains only binary data (\ie\ sequence of random bytes)
and the user should avoid opening such files directly, for example, with a text
editor. Moreover, \MMA\ is not able to import the provided
files directlly using \lstinline{Import} function and the command
\begin{lstlisting}
Import["sample_data.bin"]
\end{lstlisting}
will result in an error
\begin{lstlisting}
Import::infer: Cannot infer format of file sample_data.bin.
\end{lstlisting}
as the system is not able to provide a meaningful representation of the data
in the file. For this reason package \RandFile\ provides a level of
abstraction allowing the user to avoid dealing with low-level input/output
operations.

Code snipets included in the text assume that the file required to execute the
command can be located by \MMA\. The easiest way to assure this
is to place the necesary files in the same directory as the current notebook and
instruct \MMA\ to use it as a current directory by executing
\begin{lstlisting}
SetDirectory[NotebookDirectory[]]
\end{lstlisting}
Note that this assumes that the current notebook is already saved on the disk.

%%%%%%%%%%%%%%%%%%%%%%%%%%%%%%%%%%%%%%%%%%%%%%%%%%%%%%%%%%%%%%%%%%%%%%%%%%%%%%%%
\subsection{Installation}\label{sec:pkg-install}
%%%%%%%%%%%%%%%%%%%%%%%%%%%%%%%%%%%%%%%%%%%%%%%%%%%%%%%%%%%%%%%%%%%%%%%%%%%%%%%%
Package installation requires only to copy the \lstinline{RandFile.m} file into an
appropriate directory. As \MMA\ is supported currently on three
families of operating systems, the user should place the file in the directory
depending on the running system.

\begin{itemize}
	\item For GNU/Linux family of operating systems in
	\lstinline{.Mathematica/Applications} directory located in the user's home
	directory.
	\item For Mac OS X 10.6 and above in \lstinline{Library/Mathematica/Applications}
	directory located in the user's home directory.
	\item For Microsoft Windows XP in
\begin{lstlisting}
C:\Documents and Settings\username\Application Data\Mathematica\Applications
\end{lstlisting}
and for later versions of MS Windows in 
\begin{lstlisting}
C:\Users\username\AppData\Roaming\Mathematica\Applications
\end{lstlisting}
directory with \lstinline{username} replaced with the appropriate user name.
\end{itemize}

One can check the value of the default directory used in \MMA\
by inspecting the value of \lstinline[mathescape=false]{$UserBaseDirectory} variable.

Alternatively the user can put \lstinline{RandFile.m} file in any other location
listed in the \lstinline[mathescape=false]{$Path} variable of the \MMA\ system.

On UNIX-based operating systems (Mac OS X and GNU/Linux) the user can also use a
symbolic link in order to instal the package without the need for copying the
actual file. To do this one needs to issue the following command
\lstinline{ln -s /path/to/file/RandFile.m}
in the appropriate directory \lstinline{Library/Mathematica/Applications} for Max OS X
or \lstinline{.Mathematica/Applications} for GNU/Linux.

%%%%%%%%%%%%%%%%%%%%%%%%%%%%%%%%%%%%%%%%%%%%%%%%%%%%%%%%%%%%%%%%%%%%%%%%%%%%%%%%
\subsection{Loading and defining the random source}
%%%%%%%%%%%%%%%%%%%%%%%%%%%%%%%%%%%%%%%%%%%%%%%%%%%%%%%%%%%%%%%%%%%%%%%%%%%%%%%%

In order to use the functionality provided by the package, one has to load the
package and provide information about the source of randomness to be used by the
package. The first step can be achieved by executing
\begin{lstlisting}
<<RandFile`
\end{lstlisting}
\MMA\ command. One should note that \MMA\ is
case-sensitive and the requested package name must correspond with the file name
containing it.

As the result the user should see short information about the loaded package 
\begin{lstlisting}
Package RandFile version 0.0.18 (last modification: 24/06/2014).
\end{lstlisting}
and some additional directions concering its usage. The usage information
contains instructions about using the external source of randomess.

%%%%%%%%%%%%%%%%%%%%%%%%%%%%%%%%%%%%%%%%%%%%%%%%%%%%%%%%%%%%%%%%%%%%%%%%%%%%%%%%
\subsection{Using source of random data}
%%%%%%%%%%%%%%%%%%%%%%%%%%%%%%%%%%%%%%%%%%%%%%%%%%%%%%%%%%%%%%%%%%%%%%%%%%%%%%%%

The crucial step during the usage of \lstinline{RandFile} package is to provide the source of
random data used by the functions implemented in the package. The user should place
the file with random data in one of the directories listed in \lstinline[mathescape=false]{$Path}
global variable. The value of this variable can be displayed by issuing
\lstset{mathescape=false}
\begin{lstlisting}
$Path
\end{lstlisting}
\lstset{mathescape=true}%
command and it depends on the operating system and on the enviroment
settings of the user.

In order to install the files with random data one should follow the rules for
the instalation of the package file \lstinline{RandFile.m} as described in
Section~\ref{sec:pkg-install}. After the file with random data is placed in the
proper directory, one can use it as described in the next section. In order to
choose the source of random data one has to use \lstinline{SetTrueRandomDataFile}
function. This will set a global variable used by the package in order to
control the source of random data. During the \MMA\ session
the current value of this global variable can be displayed by issuing
\lstinline{SetTrueRandomDataFile[]} command.

For the purpose of testing the package distribution contains file
\lstinline{sample_data.bin} obtained from \cite{qrng-de} service. The file contains
1MB of random data and should provide sufficient data for testing package
functionality and performng simple experiments. When dealing with
simulations requiring larger amounts of randomness, the user is advised to obtain
true random data from \cite{qrng-de}.

One should note that \lstinline{.bin} file extension used by the data files containing
random data has been chosen in order to indicate that these files contain 
unstructured streams of binary data only. As such they are not suitable for
opening by external applications, eg. a text editor. It is also impossible to
import these files directly into Mathematica as the system cannot understand
them. 

%%%%%%%%%%%%%%%%%%%%%%%%%%%%%%%%%%%%%%%%%%%%%%%%%%%%%%%%%%%%%%%%%%%%%%%%%%%%%%%%
\subsection{Simple example}
%%%%%%%%%%%%%%%%%%%%%%%%%%%%%%%%%%%%%%%%%%%%%%%%%%%%%%%%%%%%%%%%%%%%%%%%%%%%%%%%
In order to illustrate the basic funcionality we provide a simple example of
package usage.

As usual with the external packages, the first step is to load the package by
executing 
\begin{lstlisting}
<<RandFile`
\end{lstlisting} 
or equivalently 
\begin{lstlisting}
Needs["RandFile`"]
\end{lstlisting}
and one should note the \`{} sign at the end of the package name.

After this the user has to set the source of random numbers to be used during
the session. This can be achieved by executing
\begin{lstlisting}
SetTrueRandomDataFile["sample_data.bin"] 
\end{lstlisting}
and one should note this file can be used as an argument for this command only
once during the session. However, the source of random data can be changed
during the session and the user can use \lstinline{SetTrueRandomDataFile} with a
different argument during the same session.

The above steps enable the usage of the package funcionality. For example, in
order to obtain a random integer number from $\{0,1\}$ one has to execute
\begin{lstlisting}
TrueRandomInteger[] 
\end{lstlisting}
and to sample uniformly from the interval $[0,1]$ one can use
\begin{lstlisting}
TrueRandomReal[] 
\end{lstlisting}
and by issuing 
\begin{lstlisting}
TrueRandomChoice[{a, b, c, d}]
\end{lstlisting}
one obtains a true random choice of one of $\{a, b, c, d\}$.

In general the interface provided by \RandFile\ package aims at mimicking
the standard \MMA\ interface for generating random numbers.
This allows the user to seamlessly substitute the standard functions used in
\MMA\ programs with the functions provided by the package.

A basic example with elementary usage of the random numbers interface
provided by the package can be found in \lstinline{RandFile_BasicUsage.nb} notebook
distributed with the package.

The detailed explanation of the package functionality can be found in
Appendix~\ref{app:pkg-desc}. In particular Section \ref{app:extending} of the
appendix provides an example of the procedure required for extending the package
functionality with new functions to sample from non-uniform
distributions.

%%%%%%%%%%%%%%%%%%%%%%%%%%%%%%%%%%%%%%%%%%%%%%%%%%%%%%%%%%%%%%%%%%%%%%%%%%%%%%%%
\section{Comparison with other sources of randomness}\label{sec:comparison}
%%%%%%%%%%%%%%%%%%%%%%%%%%%%%%%%%%%%%%%%%%%%%%%%%%%%%%%%%%%%%%%%%%%%%%%%%%%%%%%%
The main source of random numbers provided by \MMA\ is the built-in
pseudo-random number generator. One of the alternative sources of randomness is
provided by the \pkg{TRQS} package \cite{miszczak12generating,miszczak13employing},
which allows accessing the quantum true random number generators, using an
off-line \cite{quantis-support}, as well as an on-line
interface~\cite{qrng-de}.

%%%%%%%%%%%%%%%%%%%%%%%%%%%%%%%%%%%%%%%%%%%%%%%%%%%%%%%%%%%%%%%%%%%%%%%%%%%%%%%%
\subsection{User interface}
%%%%%%%%%%%%%%%%%%%%%%%%%%%%%%%%%%%%%%%%%%%%%%%%%%%%%%%%%%%%%%%%%%%%%%%%%%%%%%%%
Package \RandFile\ was designed for accessing sources of true random data in the
form suitable for \MMA\ users. For this reason the functions provided by the
package follow the convention used in the standard functions as close as
possible. The main differences between the standard functions and the functions
implemented by the package are caused by the need for assigning and managing the
access to the file with random data.

\begin{table}[ht!]
    \centering
     \begin{tabular}{|c|c|}
        \hline
        Standard \MMA\ pattern & \RandFile\ implementation \\
        \hline\hline
        --- & \lstinline/SetTrueRandomDataFile[fName]/ \\
        \hline
        \lstinline/BlockRandom[]/ & \lstinline/BlockTrueRandom[]/ \\
        \hline
        \lstinline/SeedRandom[]/ & \lstinline/TrueRandomSequence[]/ \\ 
        \hline
        \lstinline/SeedRandom[n]/ & \lstinline/TrueRandomSequence[n]/ \\ 
        \hline 
        --- & \lstinline/SetTrueRandomDataFile[]/ \\ 
        \hline
        --- & \lstinline/SetMaxTrueRandomSequenceLength[]/ \\ 
        \hline
        \lstinline/RandomInteger[]/ & \lstinline/TrueRandomInteger[]/ \\
        \hline 
        \lstinline/RandomInteger[{a,b}]/ & \lstinline/TrueRandomInteger[{a,b}]/ \\
        \hline
        \lstinline/RandomInteger[n]/ & \lstinline/TrueRandomInteger[n]/ \\
        \hline
        \lstinline/RandomInteger[n,k]/ & \lstinline/TrueRandomInteger[n,k]/ \\
        \hline
        \lstinline/RandomInteger[{a,b}, n]/ & \lstinline/TrueRandomInteger[{a,b}, n]/ \\
        \hline
        \lstinline/RandomInteger[{a,b}, $\SubList{d}{k}$]/ & \lstinline/TrueRandomInteger[{a,b}, $\SubList{d}{k}$]/ \\
        \hline
        --- & \lstinline/TrueRandomInteger[fName]/ \\
        \hline
        --- & \lstinline/TrueRandomInteger[n,fName]/ \\
        \hline
        --- & \lstinline/TrueRandomInteger[n,k,fName]/ \\
        \hline
        --- & \lstinline/TrueRandomInteger[n,$\SubList{d}{k}$,fName]/ \\
        \hline
%        \lstinline{RandomInteger[dist]} & \lstinline{TrueRandomInteger[dist]} \\
%        \hline
%        \lstinline{RandomReal[]} & \lstinline{TrueRandomReal[]}\\
%        \hline
%        \lstinline{RandomReal[\{a,b\}]} & \lstinline{TrueRandomReal[\{a,b\}]}\\
%        \hline
%        \lstinline{RandomReal[b]} & \lstinline{TrueRandomReal[b]}\\
%        \hline
%        \lstinline{RandomReal[\{a,b\}, n]} & \lstinline{TrueRandomReal[\{a,b\},n]}\\
%        \hline
%        \lstinline{RandomReal[dist]} & \lstinline{TrueRandomReal[dist]} \\
%        \hline
%        \lstinline{RandomComplex[]} & \lstinline{TrueRandomComplex[]}\\
%        \hline
%        \lstinline{RandomComplex[\{a,b\}]} & \lstinline{TrueRandomComplex[\{a,b\}]}\\
%        \hline
%        \lstinline{RandomComplex[b]} & \lstinline{TrueRandomComplex[b]}\\
%        \hline
%        \lstinline{RandomComplex[\{a,b\}, n]} & \lstinline{TrueRandomComplex[\{a,b\},n]}\\
%        \hline
    \end{tabular}
    \caption{Comparison of standard functions provided by \MMA\
    with their counterparts implemented in \RandFile\ package. In the above
    \lstinline{fName} refers to a string with the path to the file with random data.
    The parameters used in functions implemented in \RandFile\\ follow the
    same pattern as the standard \MMA\ functions.}
    \label{tab:rand-file-funcs}
\end{table}

Table~\ref{tab:rand-file-funcs} provides a comparison of patterns defined by
\MMA\ with the patterns used by the package for the purpose of
producing numbers in \lstinline{Integer}, \lstinline{Real} and \lstinline{Complex} domains. As
one can see, the functions provided by the package follow the standard interface
in most of the cases. This allows the integration of the package with the
existing \MMA\ programs without any significant changes in the
existing code.

The main difference in the user interface provided by \RandFile\ package in
comparison to the standard functions is that all functions can be used with an
optional last argument for assigning a file with random data used during the
call. This allows the control of the source of randomness and thanks to this
using disjoint sources during the subsequent calls. This functionality
allows, when required, using different files for producing non-overlapping
sequences of true random numbers. One should note, however, that this can be
also achieved by using \lstinline{TrueRandomSequence} function.

Among other noticeable differences one can point out the fact that the
counterpart of \lstinline{SeedRandom} function provided by the package is named
\lstinline{TrueRandomSequence}. This is motivated by the requirement of providing the
non-overlapping sequence of random numbers while using the
\lstinline{TrueRandomSequence} function.

%%%%%%%%%%%%%%%%%%%%%%%%%%%%%%%%%%%%%%%%%%%%%%%%%%%%%%%%%%%%%%%%%%%%%%%%%%%%%%%%
\subsection{Package efficiency}
%%%%%%%%%%%%%%%%%%%%%%%%%%%%%%%%%%%%%%%%%%%%%%%%%%%%%%%%%%%%%%%%%%%%%%%%%%%%%%%%
One of the important characteristics of the random number generators is their
speed. This characteristic is especially important in simulations requiring
large amounts of random data.

\begin{figure}[ht!]
    \centering
    \includegraphics[scale=0.85]{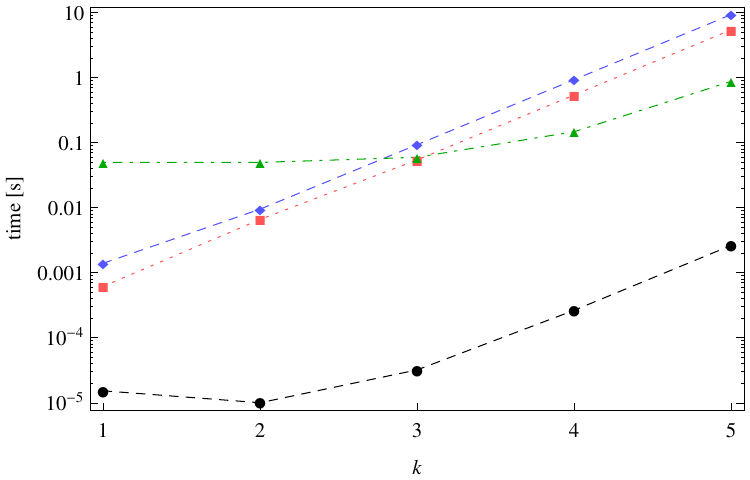}
    \caption{Comparison of the time required to generate a sample of $10^k$ real
    numbers for $k=1,\dots,5$ using pseudo-random generator provided by
    \MMA\\ (black circles), \RandFile\ using standard hard
    drive (blue diamonds), \RandFile\ using SSD hard drive (red squares) and
    the QRNG on-line generator via \pkg{TRQS} package using QRNG service (green
    triangles). The plots are based on the \lstinline{AbsoluteTiming} function and
    the results were averaged over 5 experiments.}
    \label{fig:plot-timings}
\end{figure}

Speed comparison of three methods for producing random numbers in
\MMA\ is presented in Figure~\ref{fig:plot-timings}. One can
notice that for a sample of the size larger than $10^3$ the QRNG on-line service
is able to provide random numbers with a higher speed than the \RandFile\
package. This is due to the fact that \lstinline{libQRNG} library used to access this
service allows downloading data in chunks, while \MMA\
functions used to read random data from a file can read during a call one byte
only. This is the case for \lstinline{BinaryRead}, as well as \lstinline{BinaryReadList}.

One can also notice that the use of SSD device, as it allows for the fastest
read-out of data, increases the speed of random number generation. This
illustrates the fact that the overhead of the read operations provided a
significant contribution to total speed of the number generation.

%%%%%%%%%%%%%%%%%%%%%%%%%%%%%%%%%%%%%%%%%%%%%%%%%%%%%%%%%%%%%%%%%%%%%%%%%%%%%%%%
\section{Usage examples}\label{sec:examples}
%%%%%%%%%%%%%%%%%%%%%%%%%%%%%%%%%%%%%%%%%%%%%%%%%%%%%%%%%%%%%%%%%%%%%%%%%%%%%%%%
In this section we provide some examples of the package functionality. As the
main feature of the described package is the ability to harness file-based
sources of random data, the package allows the actual manipulation of the
quality of the provided randomness. Thanks to this, the user is able to observe
the influence of quality of the provided data on the outcomes of the conducted
simulations.

The assessment of the quality of the randomness sources is a challenging task
\cite{afflerbach90criteria}. In the following, however, we use the simple
methodology implemented in the \pkg{ENT} program~\cite{ent} and we will focus on
the mean and the entropy of the input files used in the experiments.

%%%%%%%%%%%%%%%%%%%%%%%%%%%%%%%%%%%%%%%%%%%%%%%%%%%%%%%%%%%%%%%%%%%%%%%%%%%%%%%%
\subsection{Example 1: Value of pi}
%%%%%%%%%%%%%%%%%%%%%%%%%%%%%%%%%%%%%%%%%%%%%%%%%%%%%%%%%%%%%%%%%%%%%%%%%%%%%%%%
As the first example we use the Monte Carlo method for calculating the value of
$\pi$. The error in calculating this value can be used as a randomness
indicator~\cite{ent}. 

Using the functions provided by \RandFile\ package one can easily implement
a method for calculating the value of $\pi$ using a given input file with random
data. \MMA\ function used in this example is presented
in the following code.

\begin{lstlisting}
TrueRandomPi[fName_String] := With[{samplePoints = 5000},
  SetTrueRandomDataFile[fName];
  data = Table[TrueRandomReal[{0, 1}], {i, samplePoints}, {j, 2}];
  CloseTrueRandomDataFile[];
  
  dataCircle = 
    Select[data, (1/2 - #[[1]])^2 + (1/2 - #[[2]])^2 <= 1/4 &];
  complement = Complement[data, dataCircle];
  If[Length[complement] == 0, complement = {{1, 1}}];  
  4*Length[dataCircle]/samplePoints
];
\end{lstlisting}

\begin{figure}[ht!]
    \centering
    \includegraphics[scale=0.85]{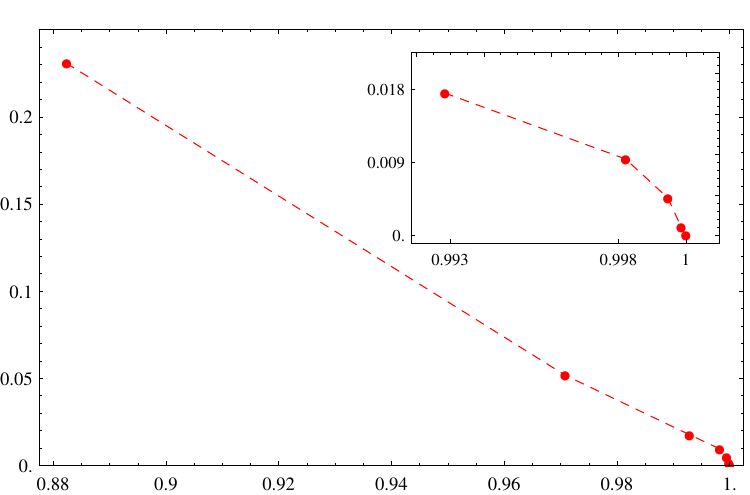}
    \caption{Relative error of calculating the value of $\pi$ using the Monte
    Carlo method as a function of entropy of the input file used to generate
    samples. The value of $\pi$ has been calculated using the function
    \lstinline{TrueRandomPi} for 7 sample files with different entropies. For the
    values of entropies close to 1, the values of the relative error are plotted
    in the inset.}
    \label{fig:relative-error-pi}
\end{figure}

The relative error of calculating the value of $\pi$ using a random file is
presented in Figure~\ref{fig:relative-error-pi}. Input files used in the
experiment have been generated using \lstinline{TrueRandomChoice} function with
slightly biased weights. The entropy and the mean of the used files treated as
sequences of bits are presented in Table~\ref{tab:parameters-of-files-pi}.

\begin{table}
    \centering
    \begin{tabular}{|l||c|c|c|c|c|c|c|}
    \hline
        bias & 0.2 & 0.1 & 0.05 & 0.025 & 0.0125 & 0.00625 & 0.0 \\\hline
        mean & 0.6991 & 0.6002 & 0.5497 & 0.5249 & 0.5135 & 0.5069 & 0.5006 \\\hline
        entropy & 0.8823 & 0.9708 & 0.9928 & 0.9982 & 0.9994 & 0.9998 & 0.9999 \\\hline
    \end{tabular}
    \caption{Parameters of the sample files used for calculating the
    value of $\pi$ using Monte Carlo method. The entropy and the mean are
    calculated for the bit sequences obtained by using
    \lstinline{TrueRandomChoice} function with weights biased towards 1. Each
    file contains 80KB of data.}
    \label{tab:parameters-of-files-pi}
\end{table}

\begin{figure}[htp!]
    \centering
    \subfigure[$\pi\approx 2.4168$]{
        \includegraphics[width=0.225\textwidth]{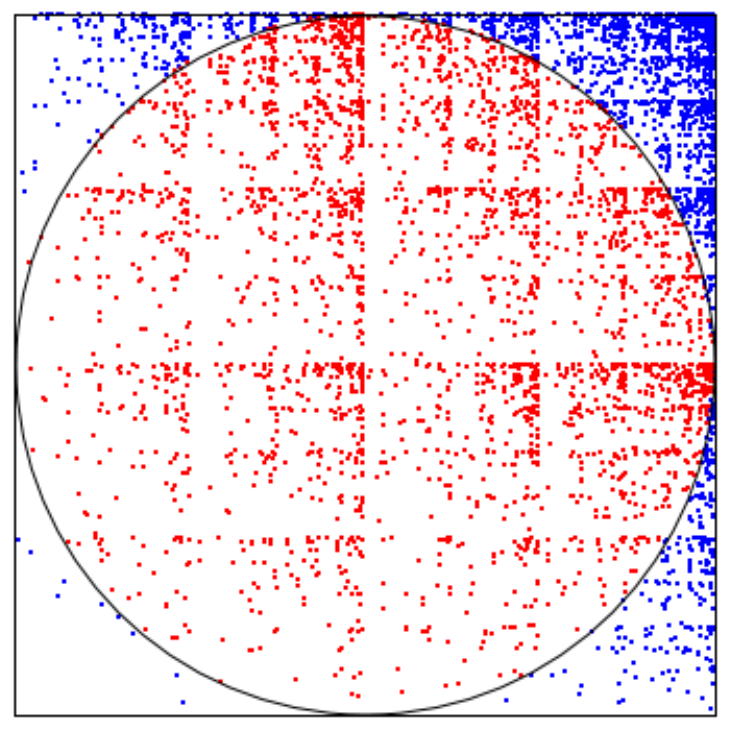}
        \label{fig:pi-hits-sample1}
    }
    \subfigure[$\pi\approx 2.9792$]{
        \includegraphics[width=0.225\textwidth]{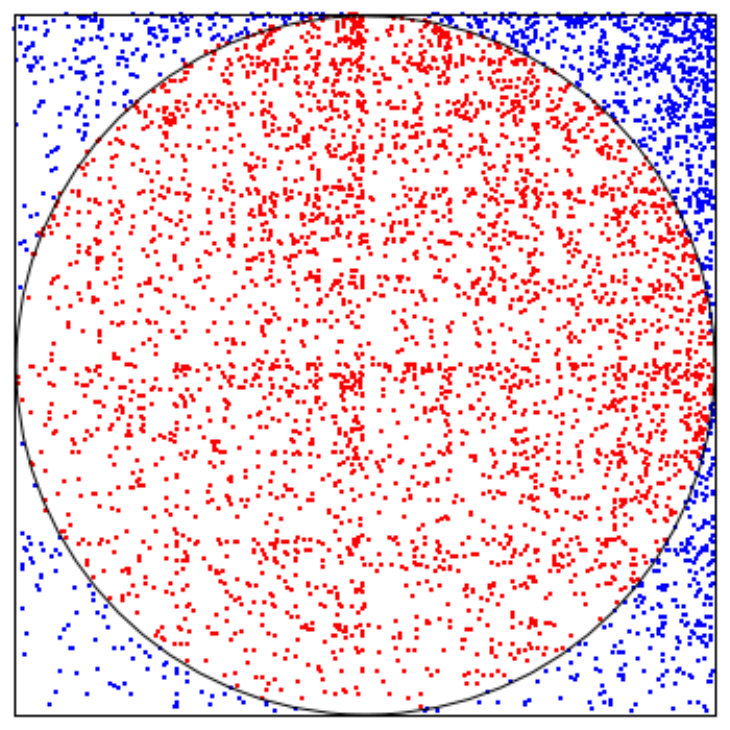}
        \label{fig:pi-hits-sample2}
    }
    \subfigure[$\pi\approx 3.1120$]{
        \includegraphics[width=0.225\textwidth]{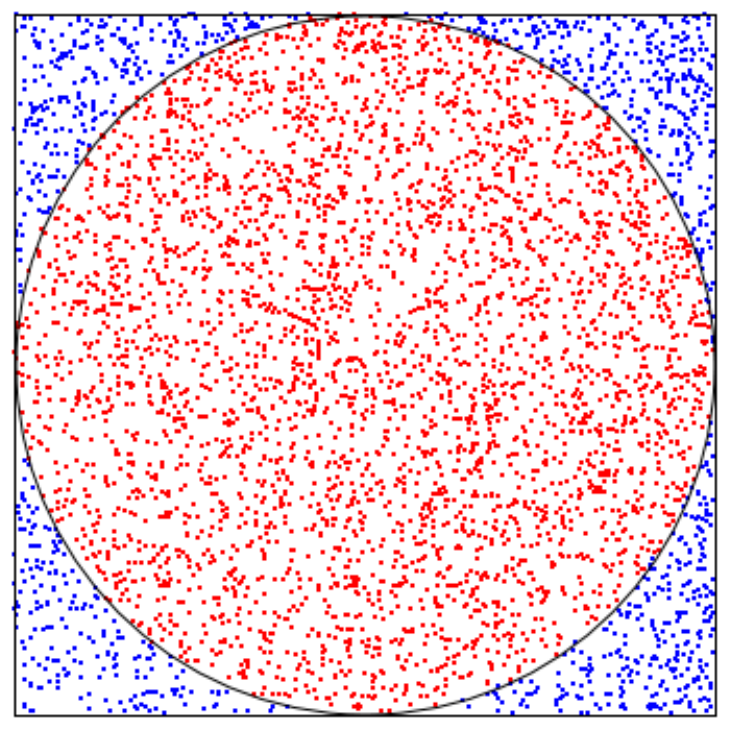}
        \label{fig:pi-hits-sample4}
    }
    \subfigure[$\pi\approx 3.1416$]{
        \includegraphics[width=0.225\textwidth]{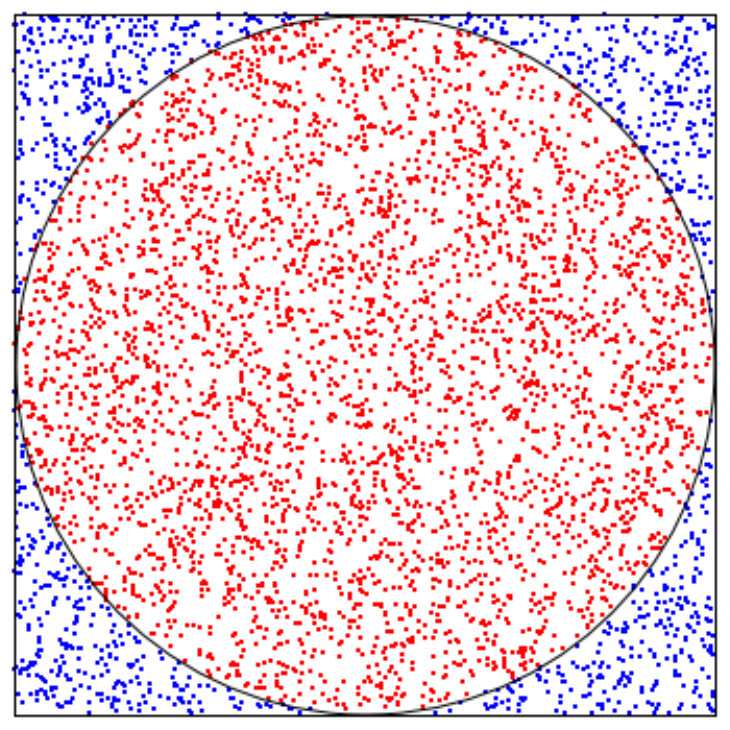}
        \label{fig:pi-hits-sample7}
    }
    \caption{Representation of the samples used to calculate the value of $\pi$
    for samples with bias towards 1 equal 0.2 \subref{fig:pi-hits-sample1}, 0.1
    \subref{fig:pi-hits-sample2}, 0.025 \subref{fig:pi-hits-sample4} and
    unbiased \subref{fig:pi-hits-sample7}. See
    Table~\ref{tab:parameters-of-files-pi} for the parameters of the
    corresponding input files. Each sample contains 5000 points.}
    \label{fig:pi-hits}
\end{figure}

The samples obtained for selected input files are depicted in
Figure~\ref{fig:pi-hits}. One can notice a clear pattern visible for the files
with largely biased distribution of bits -- Figures~\ref{fig:pi-hits-sample1}
and \ref{fig:pi-hits-sample2}.

%%%%%%%%%%%%%%%%%%%%%%%%%%%%%%%%%%%%%%%%%%%%%%%%%%%%%%%%%%%%%%%%%%%%%%%%%%%%%%%%
\subsection{Example 2: Double slit experiment}
%%%%%%%%%%%%%%%%%%%%%%%%%%%%%%%%%%%%%%%%%%%%%%%%%%%%%%%%%%%%%%%%%%%%%%%%%%%%%%%%
To illustrate the influence of the bias in the source of random data on a
physical experiment, we illustrate the usage of \RandFile\ package for the
purpose of simulating double slit experiment.

\begin{table}
    \centering
    \begin{tabular}{|l||c|c|c|c|c|c|c|}
    \hline
        bias & 0.2 & 0.1 & 0.05 & 0.025 & 0.0125 & 0.00625 & 0.0 \\\hline
        mean & 0.7000 & 0.6001 & 0.5501 & 0.5251 & 0.5126 & 0.5064 & 0.4999 \\\hline
        entropy & 0.8811 & 0.9708 & 0.9927 & 0.9981 & 0.9995 & 0.9998 & 0.9999 \\\hline
    \end{tabular}
    \caption{Parameters of the sample files used for visualizing double slit
    experiment. The entropy and the mean are calculated for the bit sequences
    obtained by using \lstinline{TrueRandomChoice} function with weights biased
    toward 1. Each file contains 2MB of data.}
    \label{tab:parameters-of-files-dse}
\end{table}

\begin{figure}[hb!]
    \centering
    \subfigure[bias = $0.2$]{
        \includegraphics[width=0.45\textwidth]{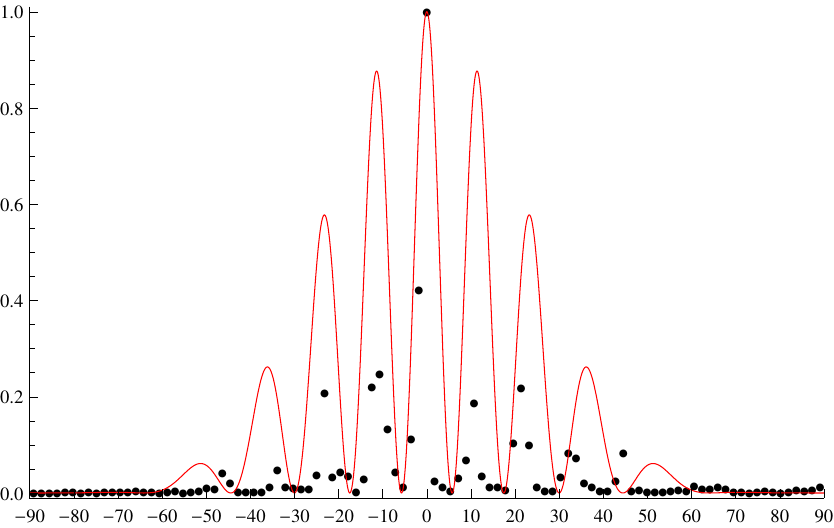}
        \label{fig:dse-sample1}
    }
    \subfigure[bias = $0.1$]{
        \includegraphics[width=0.45\textwidth]{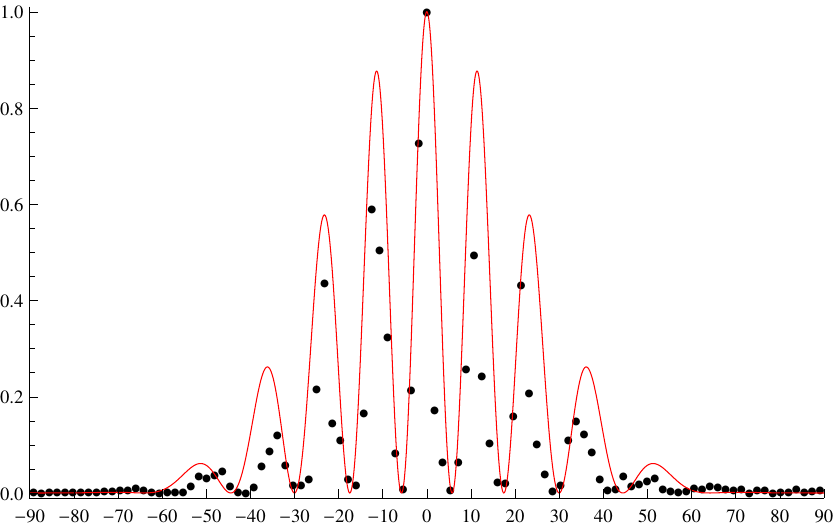}
        \label{fig:dse-sample2}
    }\\
    \subfigure[bias = $0.025$]{
        \includegraphics[width=0.45\textwidth]{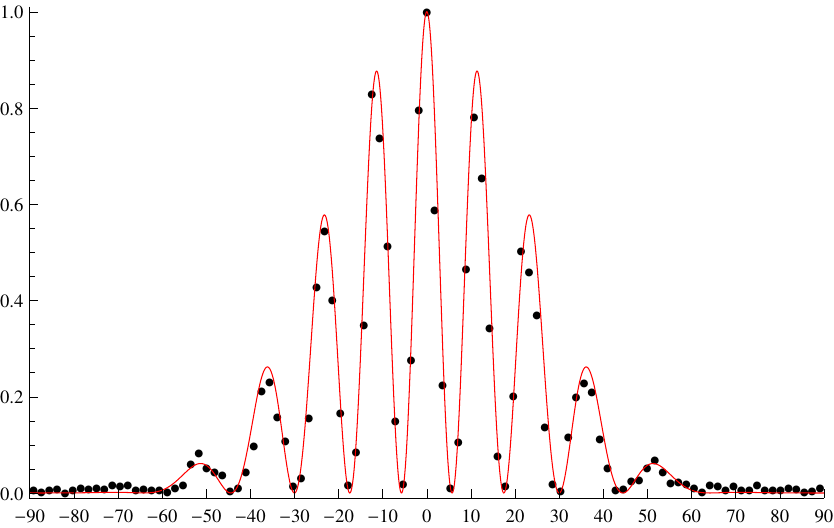}
        \label{fig:dse-sample4}
    }
    \subfigure[unbiased]{
        \includegraphics[width=0.45\textwidth]{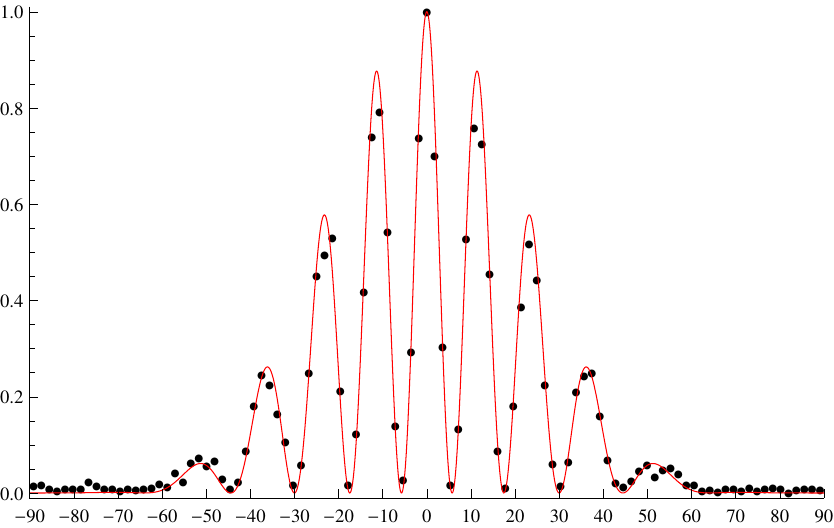}
        \label{fig:dse-sample7}
    }
    \caption{Relative intensity for angle $\theta\in(-\frac{\pi}{2},
    \frac{\pi}{2})$ calculated for simulation with $1.5 \times10^5$ events.
    Solid red line represents the theoretical value.}
    \label{fig:dse}
\end{figure}

The parameters of the sample files used in this example are provided in
Table~\ref{tab:parameters-of-files-dse}. Simulation results are presented in
Figure~\ref{fig:dse}. Each plot represents relative intensities obtained for 101
detectors and using $1.5 \times10^5$ events.

The plots presented in Figure~\ref{fig:dse} visualize the influence of the
quality of random numbers on the simulation results. In particular, for the
input files with biased distribution of bits -- Figures~\ref{fig:dse-sample1}
and \ref{fig:dse-sample2} -- the simulation results are significantly different
from the theoretical value.  

%%%%%%%%%%%%%%%%%%%%%%%%%%%%%%%%%%%%%%%%%%%%%%%%%%%%%%%%%%%%%%%%%%%%%%%%%%%%%%%
\section{Concluding remarks}\label{sec:final}
%%%%%%%%%%%%%%%%%%%%%%%%%%%%%%%%%%%%%%%%%%%%%%%%%%%%%%%%%%%%%%%%%%%%%%%%%%%%%%%%
The main contribution of this paper is the description of a package for
\MMA\\ computing system which allows the production of random
numbers using random data stored in files. This functionality enables the employment of
high-quality true random numbers in simulations. At the same time it
allows the observation of the influence of the quality of random data provided
on the outcomes of the numerical experiments.

The presented package uses local files for generating random numbers and it
provides higher reliability in accessing random numbers during the simulation
compared to the on-line true random number generators. Moreover, there is no need for any special
hardware during the simulation and the data used can be obtained using various
methods, including quantum true random number
generators~\cite{wahl11ultrafast,quantis-support} or methods based on other
physical phenomena~\cite{hotbits,random.org}.

The comparison of the efficiency of \RandFile\ with the standard
pseudo-random generator provided by \MMA\ shows that the former
is significantly faster. This can have a significant influence on the performance of
the simulations. On the other hand, the use of truly random data may be crucial
for the correctness of the simulation results. However, the presented package
does not require any external libraries and is based on \MMA\
programming languages only.

%%%%%%%%%%%%%%%%%%%%%%%%%%%%%%%%%%%%%%%%%%%%%%%%%%%%%%%%%%%%%%%%%%%%%%%%%%%%%%%%
\paragraph*{Acknowledgements}
%%%%%%%%%%%%%%%%%%%%%%%%%%%%%%%%%%%%%%%%%%%%%%%%%%%%%%%%%%%%%%%%%%%%%%%%%%%%%%%%
The presented work was possible thanks to many interesting discussions with
R.~Heinen. 
JAM would like to acknowledge financial support from the Polish National Science
Centre under the research project DEC-2011/03/D/ST6/00413.

%%%%%%%%%%%%%%%%%%%%%%%%%%%%%%%%%%%%%%%%%%%%%%%%%%%%%%%%%%%%%%%%%%%%%%%%%%%%%%%%
%\bibliographystyle{plain}
%\bibliography{rand_file}

\begin{thebibliography}{10}

\bibitem{afflerbach90criteria}
L.~Afflerbach.
\newblock Criteria for the assessment of random number generators.
\newblock {\em Journal of Computational and Applied Mathematics}, 31(1):3 --
  10, 1990.

\bibitem{bauke04pseudo}
H.~Bauke and S.~Mertens.
\newblock Pseudo random coins show more heads than tails.
\newblock {\em Journal of Statistical Physics}, 114(3-4):1149--1169, 2004.
\newblock arXiv:cond-mat/0307138.

\bibitem{box58note}
G.E.P. Box and M.E. Muller.
\newblock A note on the generation of random normal deviates.
\newblock {\em The Annals of Mathematical Statistics}, 29(2):610--611, 1958.

\bibitem{devroye86nonuniform}
L.~Devroye.
\newblock {\em Non-Uniform Random Variate Generation}.
\newblock Springer-Verlag, New York, 1986.

\bibitem{ferrenberg92montecarlo}
A.M. Ferrenberg, D.P. Landau, and Y.J. Wong.
\newblock {Monte Carlo simulations: Hidden Errors From ``Good'' Random Number
  Generators}.
\newblock {\em Physical Review Letters}, 69(23):3382--3384, 1992.

\bibitem{fishman}
G.S. Fishman.
\newblock {\em {Monte Carlo. Concepts, Algorithms and Applications}}.
\newblock Springer Series in Operations Research. Springer-Verlag, New York,
  U.S.A., 1995.

\bibitem{random.org}
M.~Haahr.
\newblock {Random.org} -- {True Random Number Service}, 1998.

\bibitem{hayes93wheel}
B.~Hayes.
\newblock The wheel of fortune.
\newblock {\em American Scientist}, 81(2):114--118, March-April 1993.

\bibitem{quantis-support}
{ID Quantique SA}.
\newblock {\em {Quantis} -- {True Random Number Generator Exploiting Quantum
  Physics}}, 2013.
\newblock Product web page
  \url{http://www.idquantique.com/random-number-generators/products.html}.

\bibitem{kanter10optical}
I.~Kanter, Y.~Aviad, I.~Reidler, E.~Cohen, and M.~Rosenbluh.
\newblock An optical ultrafast random bit generator.
\newblock {\em Nature Photonics}, 4:58--61, 2010.

\bibitem{knuth-vol2}
D.~Knuth.
\newblock {\em The Art of Computer Programming. Vol. 2: Seminumerical
  Algorithms}.
\newblock 3rd edition, 1997.

\bibitem{metropolis87beginning}
N.~Metropolis.
\newblock {The Beginning of the Monte Carlo Method}.
\newblock {\em Los Alamos Science}, (Special Issue 1987):125--130, 1987.
\newblock Special Issue dedicated to Stanis{\l}aw Ulam.

\bibitem{miszczak12generating}
J.A. Miszczak.
\newblock {Generating and Using Truly Random Quantum States in Mathematica}.
\newblock {\em Computer Physics Communications}, 183(1):118--124, 2012.
\newblock arXiv:1102.4598.

\bibitem{miszczak13employing}
J.A. Miszczak.
\newblock {Employing Online Quantum Random Number Generators for Generating
  Truly Random Quantum States in Mathematica}.
\newblock {\em Computer Physics Communications}, 184(1):257--258, 2013.
\newblock arXiv:1208.3970.

\bibitem{randfile-www}
J.A. Miszczak.
\newblock {\em {\pkg{RandFile} Package for \proglang{Mathematica}}}, 2013.

\bibitem{park88random}
S.K. Park and K.W. Miller.
\newblock {Random Number Generators: Good Ones are Hard to Find}.
\newblock {\em Communications of the ACM}, 31(10):1192--1201, 1988.

\bibitem{qrng-de}
{\em {QRNG Service}}, 2011.
\newblock Service provided by PicoQuant GmbH and the Nano-Optics groups at the
  Department of Physics of Humboldt University.

\bibitem{comqm-notes}
R.~Schmied.
\newblock {Introduction to Computational Quantum Mechanics Fall Semester 2012},
  2012.
\newblock Lecture notes available on-line from
  \url{http://atom.physik.unibas.ch/teaching/CompQM.pdf}.

\bibitem{wahl11ultrafast}
M.~Wahl, M.~Leifgen, M.~Berlin, T.~Rohlicke, H.-J. Rahn, and O.~Benson.
\newblock {An Ultrafast Quantum Random Number Generator with Provably Bounded
  Output Bias Based on Photon Arrival Time Measurements}.
\newblock {\em Applied Physics Letters}, 98(17):171105, 2011.

\bibitem{hotbits}
J.~Walker.
\newblock {\em {HotBits}: {Genuine Random Numbers, Generated by Radioactive
  Decay}}, 2006.

\bibitem{ent}
J.~Walker.
\newblock {\em {\pkg{ENT}} -- {A Pseudorandom Number Sequence Test Program}},
  2008.

\end{thebibliography}

%%%%%%%%%%%%%%%%%%%%%%%%%%%%%%%%%%%%%%%%%%%%%%%%%%%%%%%%%%%%%%%%%%%%%%%%%%%%%%%%

\appendix

%%%%%%%%%%%%%%%%%%%%%%%%%%%%%%%%%%%%%%%%%%%%%%%%%%%%%%%%%%%%%%%%%%%%%%%%%%%%%%%%
\section{Description of the package}\label{app:pkg-desc}
%%%%%%%%%%%%%%%%%%%%%%%%%%%%%%%%%%%%%%%%%%%%%%%%%%%%%%%%%%%%%%%%%%%%%%%%%%%%%%%%

%%%%%%%%%%%%%%%%%%%%%%%%%%%%%%%%%%%%%%%%%%%%%%%%%%%%%%%%%%%%%%%%%%%%%%%%%%%%%%%%
\subsection{Controlling the source of random data}
%%%%%%%%%%%%%%%%%%%%%%%%%%%%%%%%%%%%%%%%%%%%%%%%%%%%%%%%%%%%%%%%%%%%%%%%%%%%%%%%
Almost all functions provided by the package require setting a global variable
pointing to the file with random data. This can be achieved by using
\lstinline{SetTrueRandomDataFile} function. For example
\begin{lstlisting}
SetTrueRandomDataFile["sample_data.bin"]
\end{lstlisting}
assuming that \MMA\ is able to locate this file.
Alternatively, the user can provide full path to the file containing random
data, for example
\begin{lstlisting}
SetTrueRandomDataFile["C:\path\to\my_files\sample_data.bin"]
\end{lstlisting}
on Microsoft Windows operating system or
\begin{lstlisting}
SetTrueRandomDataFile["/Path/To/myFiles/sample_data.bin"]
\end{lstlisting}
on OS X, GNU/Linux or any UNIX-like operating system (such as Mac OS X).

The user should note that it is advised to use this function only once during a
session. After the file has been set, it is possible to produce random numbers
using one of the provided functions. For example, one can call
\begin{lstlisting}
TrueRandomInteger[]
\end{lstlisting}
to obtain $0$ or $1$.

Some functions provided by the package accept an extra argument which can be
used to assign the file which should be used during the call. Using this
functionality the above result can be reproduced using the command
\begin{lstlisting}
TrueRandomInteger["sample_data.bin"]
\end{lstlisting}
assuming the used file is located in one of the directories included in
\lstinline[mathescape=false]{$Path} \MMA\	 variable.

If one intends to use \lstinline{TrueRandomSequence} function they must declare
the maximal sequence length using \lstinline{SetMaxTrueRandomSequenceLength}.
Random sequences are declared by using \lstinline{TrueRandomSequence[n]}
function. The declared sequences can be displayed by calling
\lstinline{GetTrueRandomDataMarkers[]}. Once defined, the used maximal length
cannot be changed during the session.

The manipulation of the source of random data can be achieved using the
following functions.
\begin{itemize}
    \item \lstinline{SetTrueRandomDataFile[fName]} sets the global variable
    \lstinline{TrueRandomDataFile} used by the functions for generating random
    numbers. For example
    \begin{lstlisting}
    SetTrueRandomDataFile["/Path/To/myFiles/sample_data.bin"]    
    \end{lstlisting}
	on UNIX-like operating systems or
    \begin{lstlisting}
    SetTrueRandomDataFile["C:\path\to\my_files\sample_data.bin"]    
    \end{lstlisting}
	on Microsoft Windows systems.
    \item \lstinline{GetTrueRandomDataFile[]} displays the value of the global
    variable \lstinline{TrueRandomDataFile}.
    \item \lstinline{CloseTrueRandomDataFile[]} closes the file assigned by the
    global variable set using \lstinline{SetTrueRandomDataFile}.
    \item \lstinline{TrueRandomDataFile} is the global variable, defined in
    \lstinline{RandFile`Private} context, storing a string with the name of the
    file containing random data.
    \item \lstinline{TrueRandomDataFileBytesCount} is the global variable for
    storing a number of bytes in the \lstinline{TrueRandomDataFile}.
    \item \lstinline{BlockTrueRandom[ex]} evaluates expression \lstinline{ex}
    and stores the starting position in the file \lstinline{TrueRandomDataFile}.
    The position in the file is restored after the execution. Note that this
    function works only with expressions using the global variable for pointing
    to a file with random data.
    \item \lstinline{TrueRandomSequence[]} changes the current position in the
    file assigned by global variable \lstinline{TrueRandomDataFile}, so that the
    numbers generated after the execution of this function will not overlap with
    the previously generated numbers.
    \item \lstinline{TrueRandomSequence[pos]} marks the current position in the
    file with random data or, if the position \lstinline{pos} is already marked,
    returns to this position in the file.
    \item \lstinline{TrueRandomDataMarkers} is a global variable for
    accumulating positions in the file with random data used by
    \lstinline{TrueRandomSequence} function. Note that this array is sorted with
    respect to the second element.
    \item \lstinline{SetMaxTrueRandomSequenceLength[len]} declares the maximal
    length, expressed in bytes, of the true random sequence used during the
    session.
    \item \lstinline{GetMaxTrueRandomSequenceLength[]} displays the maximal
    length of the random sequence which can be used during the session.
    \item \lstinline{GetTrueRandomDataMarkers[]} displays the currently set
    markers in the data file that are used by
    \lstinline{TrueRandomSequence}.
\end{itemize}

%%%%%%%%%%%%%%%%%%%%%%%%%%%%%%%%%%%%%%%%%%%%%%%%%%%%%%%%%%%%%%%%%%%%%%%%%%%%%%%%
\subsection{Obtaining random numbers}
%%%%%%%%%%%%%%%%%%%%%%%%%%%%%%%%%%%%%%%%%%%%%%%%%%%%%%%%%%%%%%%%%%%%%%%%%%%%%%%%
Package \RandFile\ aims to provide a user interface similar to the standard
user interface for using random numbers provided by \MMA\. The
user should be able to incorporate the package into existing simulations with
minimal changes in the source code only.

%%%%%%%%%%%%%%%%%%%%%%%%%%%%%%%%%%%%%%%%%%%%%%%%%%%%%%%%%%%%%%%%%%%%%%%%%%%%%%%%
\subsubsection{Integer numbers}\label{sec:rand-file-integers}
%%%%%%%%%%%%%%%%%%%%%%%%%%%%%%%%%%%%%%%%%%%%%%%%%%%%%%%%%%%%%%%%%%%%%%%%%%%%%%%%
The package provides the following functions for producing random integer
numbers.
\begin{itemize}
    \item \lstinline/TrueRandomInteger[]/ produces an integer number in $[0,1]$.
	\item \lstinline/TrueRandomInteger[n]/ produces an integer number in $[0,n]$.
	\item \lstinline/TrueRandomInteger[n,k]/ produces \lstinline{k} integer numbers in $[0,n]$.
    \item \lstinline/TrueRandomInteger[{a,b}]/ produces an integer number in
    $[a,b)$.
    \item \lstinline/TrueRandomInteger[{a,b},k]/ produces \lstinline{k} integer
    numbers in $[a,b)$.
    \item \lstinline/TrueRandomInteger[{a,b},$\SubList{d}{k}]$/ produces a
    \SubList{d}{k}-dimensional array of integer numbers in $[a,b)$.
\end{itemize}

All of the above functions have their counterparts of the form
\begin{lstlisting}
TrueRandomInteger[params, fName]
\end{lstlisting}

For example, the command
\begin{lstlisting}
TrueRandomInteger[n]
\end{lstlisting}
will produce a table with \lstinline{n} elements being 1 or 0 using a global file with random data, while
using
\begin{lstlisting}
TrueRandomInteger[n, "sample_data.bin"]
\end{lstlisting}
will generate a similar result using data from file
\lstinline{sample_data.bin}

The user should be aware of the fact that the amount of random bytes utilized by
the functions for producing integer numbers depends on the range.

%%%%%%%%%%%%%%%%%%%%%%%%%%%%%%%%%%%%%%%%%%%%%%%%%%%%%%%%%%%%%%%%%%%%%%%%%%%%%%%%
\subsubsection{Real numbers}
%%%%%%%%%%%%%%%%%%%%%%%%%%%%%%%%%%%%%%%%%%%%%%%%%%%%%%%%%%%%%%%%%%%%%%%%%%%%%%%%
As in the case of integer numbers, the functions for producing real numbers use,
by default, the source of random numbers assigned in the global variable set by
\lstinline{SetTrueRandomDataFile}. To obtain random real numbers one can use one
of the following functions.
\begin{itemize}
    \item \lstinline/TrueRandomReal[]/ produces a 32-bit real number in $[0,1]$.
    \item \lstinline/TrueRandomReal[b]/ produces a 32-bit real number in $[0,b]$.
    \item \lstinline/TrueRandomReal[b,k]/ produces \lstinline{k} 32-bit real numbers in $[0,b]$.
    \item \lstinline/TrueRandomReal[{a,b}]/ produces a 32-bit real number in
    $[a,b]$.
    \item \lstinline/TrueRandomReal[{a,b}, k]/ produces \lstinline{k} 32-bit real
    numbers in $[a,b]$.
    \item \lstinline/rueRandomReal[{a,b}, $\SubList{d}{k}$]/ produces a
    \SubList{d}{k}-dimensional array of 32-bit real numbers in $[a,b]$.
\end{itemize}

Also in this case all of the above functions have their counterparts of the form
\begin{lstlisting}
TrueRandomReal[params, fName]
\end{lstlisting}
where \lstinline{params} refers to the standard parameters, while
\lstinline{fName} can be used to assign a file with random data used during the
execution. For example, using the command
\begin{lstlisting}
TrueRandomReal[b]
\end{lstlisting}
will produce a real number in $[0,b]$ using a global file with random data,
while using the command
\begin{lstlisting}
TrueRandomReal[b, "sample_data.bin"]
\end{lstlisting}
will generate a similar result using data from file
\lstinline{sample_data.bin}

In contrast to the functions for generating random integers, the functions
producing random real numbers require a fixed number of bytes during each call
-- by default each call requires 4 bytes and this value can be changed by
altering \lstinline{TrueRandomReal} function. This allows the anticipation of
the size of the required data files or the lenght of sequences defined by
\lstinline{SetMaxTrueRandomSequenceLength} and \lstinline{TrueRandomSequence}
functions.

%%%%%%%%%%%%%%%%%%%%%%%%%%%%%%%%%%%%%%%%%%%%%%%%%%%%%%%%%%%%%%%%%%%%%%%%%%%%%%%%
\subsubsection{Complex numbers}
%%%%%%%%%%%%%%%%%%%%%%%%%%%%%%%%%%%%%%%%%%%%%%%%%%%%%%%%%%%%%%%%%%%%%%%%%%%%%%%%
The third category of functions provided by the package for producing random
numbers can be used to obtain complex numbers. There are the following functions
in this category
\begin{itemize}
    \item \lstinline/TrueRandomComplex[]/ gives a true random complex number
    with real and imaginary parts in the range 0 to 1.
    \item \lstinline/TrueRandomComplex[{min,max}]/ gives a true random
    complex number in the rectangle with corners given by the complex numbers
    \lstinline{min} and \lstinline{max}.
    \item \lstinline/TrueRandomComplex[{min,max},n]/ gives a list of \lstinline{n}
    true random complex numbers in a given rectangle.
    \item \lstinline/TrueRandomComplex[{min,max},$\SubList{d}{k}$]/ gives
    a $\{d_1,d_2,\dots,d_k\}$-dimensional array of true random complex numbers
    in a given rectangle.
\end{itemize}

As in the case of integer and real numbers, the above functions can be called
with the third optional argument for assigning the file used to obtain random
data during the call.

%%%%%%%%%%%%%%%%%%%%%%%%%%%%%%%%%%%%%%%%%%%%%%%%%%%%%%%%%%%%%%%%%%%%%%%%%%%%%%%%
\subsubsection{Sampling functions}
%%%%%%%%%%%%%%%%%%%%%%%%%%%%%%%%%%%%%%%%%%%%%%%%%%%%%%%%%%%%%%%%%%%%%%%%%%%%%%%%
The package provides functions for obtaining weighted samples following the
convention used by \lstinline{RandomChoice} and \lstinline{RandomSample}
standard \MMA\ functions. 
\begin{itemize}
    \item \lstinline/TrueRandomChoice[$\SubList{e}{k}$]/ gives a true random
    choice of one of \SubList{e}{k}.
    \item \lstinline/TrueRandomChoice[$\SubList{e}{k}$], n]/ gives a list of
    \lstinline{n} true random choices.
    \item \lstinline/TrueRandomChoice[elist,$\SubList{d}{k}$]/ gives a
    \SubList{d}{k}-dimensional array of true random choices.
    \item \lstinline/TrueRandomChoice[$\SubList{w}{n}$ -> $\SubList{e}{n}$]/
    gives a true random choice weighted by \lstinline{wi}.
    \item \lstinline/TrueRandomChoice[$\SubList{w}{n}$ -> $\SubList{e}{n}$, k]/
    gives \lstinline{k} true random choices weighted by \lstinline{wi}.
    \item \lstinline/TrueRandomChoice[$\SubList{w}{n}$ -> $\SubList{e}{n}$ , $\SubList{d}{k}$]/
    gives a \SubList{d}{k}-dimensional array of true random choices weighted
    by \lstinline{wi}.
    \item \lstinline/TrueRandomSample[l]/ gives a true random permutation of
    the list \lstinline{l}.
    \item \lstinline/TrueRandomSample[l,n]/ gives $n$ elements from the true
    random sample of the list \lstinline{l}. Note that it is possible to take at most
    \lstinline{Length[l]} elements.
    \item \lstinline/TrueRandomSample[$\SubList{w}{n}$ -> $\SubList{e}{n}$, k]/
    gives $k$ elements from the non-uniform sample of elements \SubList{e}{n}
    with weights \SubList{w}{n}.
\end{itemize}

%%%%%%%%%%%%%%%%%%%%%%%%%%%%%%%%%%%%%%%%%%%%%%%%%%%%%%%%%%%%%%%%%%%%%%%%%%%%%%%%
\subsubsection{Deprecated Random[] interface}
%%%%%%%%%%%%%%%%%%%%%%%%%%%%%%%%%%%%%%%%%%%%%%%%%%%%%%%%%%%%%%%%%%%%%%%%%%%%%%%%
As of version 6 \MMA\ function \lstinline{Random[]} is deprecated and one is
encouraged to use functions \lstinline{RandomInteger}, \lstinline{RandomReal}
and \lstinline{RandomComplex} instead. However, for the sake of compatibility,
\RandFile\ package implements the counterpart of the standard
\lstinline{Random[]} function.
\begin{itemize}
    \item \lstinline{TrueRandom[type,range]} gives a true random number of type
    \lstinline{Integer}, \lstinline{Real} or \lstinline{Complex} in a specified
    range.
\end{itemize}

The range specification follows the convention used in \lstinline{Random[]} function
and the user is advised to consult \MMA\ documentation in this
matter.

%%%%%%%%%%%%%%%%%%%%%%%%%%%%%%%%%%%%%%%%%%%%%%%%%%%%%%%%%%%%%%%%%%%%%%%%%%%%%%%%
\subsection{Extending the package}\label{app:extending}
%%%%%%%%%%%%%%%%%%%%%%%%%%%%%%%%%%%%%%%%%%%%%%%%%%%%%%%%%%%%%%%%%%%%%%%%%%%%%%%%
The functions implemented in the package provide only a basic interface, allowing the
generation of uniformly distributed random numbers. In order to obtain non-uniform
samples from different probablity distributions, one has to extened the
functionality offered by the package. 

Package \RandFile\ delivers functions for generating samples of integer
numbers generated with \lstinline{PoissonDistribution} and real numbers generated
with \lstinline{NormalDistribution}. Below we present a particular case --
normal distribution -- and the described functions can be used as a template
for creating user's own functions.

In order to create a new function for sampling from the normal distribution, the
user can start by defining the base function, which returns one number. In the
case of the normal distribution, this functions is of the form 
\begin{lstlisting}
TrueRandomReal[dist_NormalDistribution]:=
	Mean[dist]+Sqrt[2]Sqrt[Variance[dist]] *
		InverseErf[-1+2 TrueRandomReal[{0,1}]];
\end{lstlisting}

and it is based on the inverse transform sampling
method~\cite{devroye86nonuniform}. This method can be used to obtain any
non-uniform distribution.

In the case of the normal distribution, the inverse transform sampling method is
known to be inefficient in terms of time efficiency. For this reason it can be
desirable to use Box-Muller method~\cite{box58note}. However, one should note
that Box-Muller method is less efficient in terms of random data used as it
requires two uniformly distributed random numbers to obtain one number for the
normal distribution.

In order to mimic \MMA\ interface for obtaining random numbers,
\RandFile\ package introduces two private functions, which can be used to
build functions for obtaining arrays of random numbers. The first of these
functions is
\begin{lstlisting}
CallAsTableVersion[fun_,n_Integer] := Table[fun,{n}];
\end{lstlisting}	

which for a given function \lstinline{funCall} and an integer \lstinline{n} returns a
one-dimensional table with \lstinline{n} elements.

The second function is used in order to create an array of random numbers. For the dimensions specified in a
list \lstinline{dims} one can use this function as
\begin{lstlisting}
CallAsArrayVersion[fun_,dims_List] :=
	Fold[Partition,Table[fun,{Times@@Flatten[dims]}],Reverse[dims]][[1]];}
\end{lstlisting}

Both \lstinline{CallAsTableVersion} and \lstinline{CallAsArrayVersion} functions have \lstinline{HoldAll} attribute.
\begin{lstlisting}
SetAttributes[CallAsTableVersion,HoldAll];
SetAttributes[CallAsArrayVersion,HoldAll];
\end{lstlisting}

Using the basic function for generating random numbers with the normal
distribution, the user can implement the functions for generating arrays of
random numbers as 
\begin{lstlisting}
TrueRandomReal[dist_NormalDistribution,n_Integer] := 
	CallAsTableVersion[TrueRandomReal[dist],n];
\end{lstlisting}
and
\begin{lstlisting}
TrueRandomReal[dist_NormalDistribution,dims_List] := 
	CallAsArrayVersion[TrueRandomReal[dist],dims];
\end{lstlisting}

The above functions are implemented by \RandFile\ and can be used as follows.
\begin{itemize}
    \item \lstinline/TrueRandomReal[NormalDistribution[m,s]]/
    produces 32-bit real number sampled from
    \lstinline{NormalDistribution[m,s]}.
    \item \lstinline/TrueRandomReal[NormalDistribution[m,s], n]/
    produces \lstinline{n} 32-bit real numbers sampled from
    \lstinline{NormalDistribution[m,s]}.
    \item
    \lstinline/TrueRandomReal[NormalDistribution[m,s],$\SubList{d}{k}$]/
    produces a \SubList{d}{k}-dimensional array of 32-bit real numbers sampled
    from \lstinline{NormalDistribution[m,s]}.
\end{itemize}

Additionally, \RandFile\ package provides functions for obtaining random
numbers from the normal distribution and allowing for the explicit declaration
of the used source of randomness. These functions are similar to the functions
defined above, but they accept an additional argument, which is used to specify
the file.

The base function in this case reads
\begin{lstlisting}
TrueRandomReal[dist_NormalDistribution,fName_String]:=
	Mean[dist]+Sqrt[2]Sqrt[Variance[dist]]*
		InverseErf[-1+2 TrueRandomReal[{0,1},fName]];
\end{lstlisting}
and the versions for producing arrays of numbers read
\begin{lstlisting}
TrueRandomReal[dist_NormalDistribution,n_Integer,fName_String]:=
	CallAsTableVersion[TrueRandomReal[dist,fName],n];
\end{lstlisting}
and 
\begin{lstlisting}
TrueRandomReal[dist_NormalDistribution,dims_List,fName_String]:=
	CallAsArrayVersion[TrueRandomReal[dist,fName],dims];
\end{lstlisting}
To summarize, in order to extend the functionality of the package, the user has to:
\begin{enumerate}
	\item define a basic function for obtaining one element from the required
	sample,
	\item use \lstinline{CallAsTableVersion} and \lstinline{CallAsArrayVersion} functions
	to define functions producing arrays of random numbers,
	\item optionally define functions with an explicit argument for pointing the
	file which should be used as a source of random data. 
\end{enumerate}

\end{document}